# MULTIFIELD MODEL FOR COSSERAT MEDIA

ALEKSEY A. VASILIEV, ANDREY E. MIROSHNICHENKO
AND MASSIMO RUZZENE

We construct a two-field higher-order gradient micropolar model for Cosserat media on the basis of a square lattice of elements with rotational degrees of freedom. This model includes equations of single-field higher-order gradient micropolar theory, and additional ones, which allow modelling of short wavelength phenomena. We demonstrate an example of short wavelength spatially localized static deformations in a structural system, which could not be obtained in the classical single-field framework, but which are captured by the proposed two-field model.

## 1. Introduction

Field theories are effectively used for modeling structural systems. In particular, they help to define generalized macrocharacteristics of systems, find analytical solutions by using well-developed mathematical methods, and in cases when it is impossible to make computational investigation by using effective numerical methods and packages based on artificial discretization. There are however structural effects, which are not captured through classical continuum models. This may lead to essential errors in application. The study of such effects within the framework of the field theories requires the development of generalized continuum models. One approach to the development of such models consists in the analysis and evaluation of key physical hypotheses of existing models and, further, their rejection or generalization [Lomakin 1970; Rogula 1985]. For example, in Cosserat and micropolar models [Cosserat and Cosserat 1909; Eringen 1999; Askar 1986; Maugin 1999] rotational degrees of freedom of structural elements are taken into consideration in addition to displacements, while higher derivatives of the fields are taken into account in higher-order gradient models [Triantafyllidis and Bardenhagen 1993; Fleck and Hutchinson 1997, 2001; Peerlings et al. 2001; Askes et al. 2002; Bažant and Jirásek 2002; Aifantis 2003].

For bodies with periodic microstructure, homogenized models are constructed on the basis of an elementary cell of periodicity by using a single vector function of the generalized displacements, defining the degrees of freedom of the elementary cell. Methods of obtaining continuum models from lattice models, compari-son between discrete lattice and continuum models, their advantages and applications for solving different problems have been discussed in earlier articles [Noor 1988; Triantafyllidis and Bardenhagen 1993; Pasternak and Mühlhaus 2000; Askes et al. 2002; Suiker and de Borst 2005; Pavlov et al. 2006] and monographs [Born and Huang 1954; Askar 1986; Maugin 1999]. The derivation and analysis of generalized models starting from microstructural models is one of the key approaches to develop, explore, and find practical interpretations of corresponding phenomenological theories. By using a macrocell, comprising of several elementary unit cells, and, accordingly, by increasing the number of vector fields in order to describe the deformations of the system we come to models of multifield theory [Vasiliev and Miroshnichenko 2005; Vasiliev et al. 2005], which is discussed here.

Underlying hypotheses used in the construction of the above-mentioned theories are mutually independent, complementary and can be used in various combinations. Single-field higher-order gradient micropolar models and their applications were considered in [Mühlhaus and Oka 1996; Suiker and de Borst 2005; Pavlov et al. 2006]. A nonlocal continuum Cosserat model was presented in [Pasternak and Mühlhaus 2000]. The hierarchical system of multifield micropolar models was derived in [Vasiliev and Miroshnichenko 2005]. In the present article, we derive the higher-order gradient generalization of the two-field micropolar model introduced in [Vasiliev and Miroshnichenko 2005] and compare different models in order to define possible applications of multifield theory.

The article is organized as follows. The discrete model of a square lattice of elements with rotational degrees of freedom and its single-field higher-order gradient micropolar model are presented in Sections 2 and 3. In Section 4 we derive a two-field higher-order gradient micropolar model for a Cosserat medium. In Section 5 the comparative analysis of models for a two-dimensional case is carried out by using plane wave solutions. Particular cases of models presented in Sections 2-4 for the study of one-dimensional deformations are derived in Section 6. The comparative analysis of the models in the description of spatially localized dynamic and static deformations is presented in Sections 7 and 8. In Section 9 we briefly summarize our results, outline possible fields of applications, and describe further investigations.

## 2. Cosserat lattice

We consider a Cosserat lattice, that is, a lattice whose deformations are described by displacements $u_n$, $v_n$, and by rotations $\varphi_n$ of its elements. The elements are placed at the nodes of a square lattice as shown in Figure 1a. The potential energy associated with the elastic connection of elements *m, k* has the following form

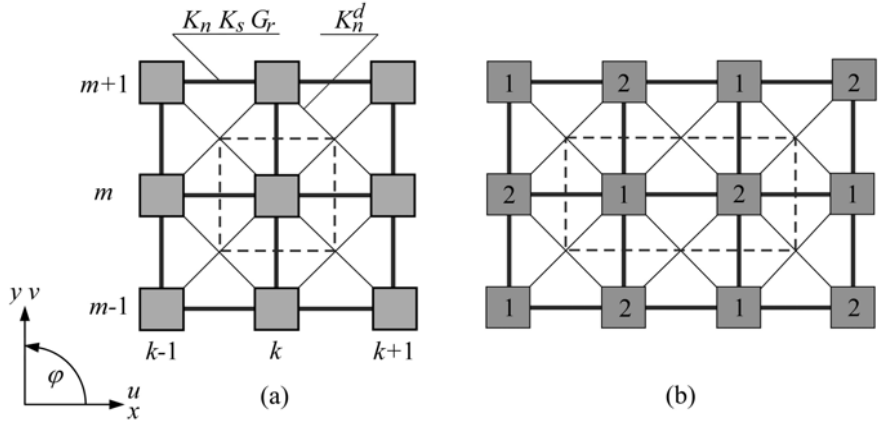

**Figure 1.** (a) An example of a square lattice consisting of elements with rotational degrees of freedoms. (b) Variant of numeration of identical elements, which is used for deriving the two-field model. The unit cell and macrocell consisting of two elements are shown by dashed lines.

$$2E_{pot}^{k,m} = K_n^{k,m}(u_m - u_k)^2 + K_s^{k,m}\left[v_m - v_k - r_{k,m}\frac{\varphi_m + \varphi_k}{2}\right]^2 + G_r^{k,m}(\varphi_m - \varphi_k)^2, \quad (1)$$

where $r_{k,m}$ is a length parameter, $K_n^{k,m}$, $K_s^{k,m}$, and $G_r^{k,m}$ characterize the stiffness of the connections in the longitudinal and transverse directions, and the resistance to rotations of elements. This form of potential energy is used in models of granular media [Limat 1988; Pasternak and Mühlhaus 2000; Suiker et al. 2001]. The potential energy of beam finite elements which is often used in lattice models of constructions and materials [Noor 1988], can be considered as a particular case of Equation (1). We use the notations $K_n$, $K_s$, and $G_r$ for the elastic constants of the axial connections, while $K_n^d$ is the axial stiffness of the diagonal connections, and we assume that $K_s^d = 0$, $G_r^d = 0$. The last assumptions mean that we assume a string type model for connections of elements in a diagonal direction. Such a model of interactions was assumed in the models of structural media proposed in [Pavlov et al. 2006]. As it was shown in [Suiker et al. 2001] it is required for the long-wave approximation for equation of motion of the square lattice to be equal to the equations of motion of the Cosserat continuum model. Very often, for example in beam lattice constructions, there are no diagonal connections at all, that is, additionally we should assume that $K_n^d = 0$. The parame-ter $r_{k,m}$ is equal to $h$ and $h\sqrt{2}$ for the axial and diagonal connections, respectively.

The expression for the kinetic energy of elements has the standard form

$$E^k_{kin} = \frac{1}{2}M\dot{u}_k^2 + \frac{1}{2}M\dot{v}_k^2 + \frac{1}{2}I\dot{\varphi}_k^2, \qquad (2)$$

where $M$ is the mass and $I$ is the moment of inertia of $k$ th element.

The equations of motion are obtained by using Lagrange's equations and have the form

$$M\ddot{u}_{k,m} = K_n\Delta_{xx}u_{k,m} + K_s\left(\Delta_{yy}u_{k,m} + \frac{1}{2}h\Delta_y\varphi_{k,m}\right) + \frac{1}{2}K_n^d\left(\Delta u_{k,m} + \Delta_{xy}v_{k,m}\right), \qquad (3)$$

$$M\ddot{v}_{k,m} = K_n\Delta_{yy}v_{k,m} + K_s\left(\Delta_{xx}v_{k,m} - \frac{1}{2}h\Delta_x\varphi_{k,m}\right) + \frac{1}{2}K_n^d\left(\Delta v_{k,m} + \Delta_{xy}u_{k,m}\right), \qquad (4)$$

$$I\ddot{\varphi}_{k,m} = \left(G_r - \frac{1}{4}K_sh^2\right)\left(\Delta_{yy}\varphi_{k,m} + \Delta_{xx}\varphi_{k,m}\right) + \frac{1}{2}K_sh\left(\Delta_x v_{k,m} - \Delta_y u_{k,m} - 4h\varphi_{k,m}\right), \qquad (5)$$

where the following notations are used

$$\Delta_x w_{k,m} = w_{k+1,m} - w_{k-1,m}, \quad \Delta_{xx}w_{k,m} = w_{k+1,m} - 2w_{k,m} + w_{k-1,m},$$
$$\Delta_y w_{k,m} = w_{k,m+1} - w_{k,m-1}, \quad \Delta_{yy}w_{k,m} = w_{k,m+1} - 2w_{k,m} + w_{k,m-1},$$
$$\Delta_{xy}w_{k,m} = w_{k+1,m+1} - w_{k+1,m-1} - w_{k-1,m+1} + w_{k-1,m-1}, \qquad (6)$$
$$\Delta w_{k,m} = w_{k+1,m+1} + w_{k+1,m-1} + w_{k-1,m+1} + w_{k-1,m-1} - 4w_{k,m}.$$

### 3. Micropolar model. Higher-order gradient micropolar model

In the micropolar model it is assumed that deformations of a discrete system can be described by using the single vector function $\{u(x,y,t), v(x,y,t), \varphi(x,y,t)\}$, which has the same components of the vector of generalized displacements $\{u_{k,m}(t), v_{k,m}(t), \varphi_{k,m}(t)\}$ of the unit cell. Accordingly, the following equalities are assumed at the nodes of the lattice

$$\{u(x,y,t), v(x,y,t), \varphi(x,y,t)\}\big|_{\substack{x=kh \\ y=mh}} = \{u_{k,m}(t), v_{k,m}(t), \varphi_{k,m}(t)\}. \qquad (7)$$

The substitution $w_{k\pm 1, m\pm 1}(t) \to w(x\pm h, y\pm h, t)$ in the finite difference equations (3)-(5) leads to the functional difference non-local equations in spatial variables. Using Taylor series expansions

$$w(x\pm h, y\pm h, t) = e^{\pm h\partial_x \pm h\partial_y}w(x,y,t) = \sum_{r=0}^{N_x}\sum_{p=0}^{N_y}\frac{(\pm h)^r}{r!}\frac{(\pm h)^p}{p!}\frac{\partial^{r+p}w(x,y,t)}{\partial x^r \partial y^p} \qquad (8)$$

gives a set of equations, which are differential with respect to spatial and temporal variables. The expansion (8) is exact and the continuum equations are exact if the Taylor series expansions include all derivatives, that is, $N_x = \infty$ and $N_y = \infty$. Truncation of the series to the $N$ th order, that is, up to $N_x + N_y \leq N$, yields the formulation of approximated continuum models.

Keeping derivatives up to the fourth order leads to the following single-field higher-order gradient micropolar model

$$Mu_{tt} = \left(K_n + K_n^d\right)h^2 u_{xx} + \left(K_s + K_n^d\right)h^2 u_{yy} + 2K_n^d h^2 v_{xy} + K_s h^2 \varphi_y$$
$$+ \frac{1}{12}\left(K_n + K_n^d\right)h^4 u_{xxxx} + \frac{1}{12}\left(K_s + K_n^d\right)h^4 u_{yyyy} + \frac{1}{2}K_n^d h^4 u_{xxyy} \quad (9)$$
$$+ \frac{1}{3}K_n^d h^4 \left(v_{xyyy} + v_{xxxy}\right) + \frac{1}{6}K_s h^4 \varphi_{yyy},$$

$$Mv_{tt} = \left(K_s + K_n^d\right)h^2 v_{xx} + \left(K_n + K_n^d\right)h^2 v_{yy} + 2K_n^d h^2 u_{xy} - K_s h^2 \varphi_x$$
$$+ \frac{1}{12}\left(K_s + K_n^d\right)h^4 v_{xxxx} + \frac{1}{12}\left(K_n + K_n^d\right)h^4 v_{yyyy} + \frac{1}{2}K_n^d h^4 v_{xxyy} \quad (10)$$
$$+ \frac{1}{3}K_n^d h^4 \left(u_{xyyy} + u_{xxxy}\right) - \frac{1}{6}K_s h^4 \varphi_{xxx},$$

$$I\varphi_{tt} = \left(G_r - \frac{1}{4}K_s h^2\right)h^2 \left(\varphi_{xx} + \varphi_{yy}\right) + K_s h^2 \left(v_x - u_y - 2\varphi\right)$$
$$+ \frac{1}{12}\left(G_r - \frac{1}{4}K_s h^2\right)h^4 \left(\varphi_{xxxx} + \varphi_{yyyy}\right) + \frac{1}{6}K_s h^4 \left(v_{xxx} - u_{yyy}\right). \quad (11)$$

The single-field model with derivatives up to the second order and its comparison with a conventional micropolar model are presented in [Suiker et al. 2001].

## 4. Two-field higher-order gradient micropolar model

The method of deriving a hierarchical system of multifield models was proposed in [Vasiliev and Miroshnichenko 2005]. In the present article, only some basic ideas will be presented, and the two-field higher-order gradient micropolar model will be obtained.

The single-field micropolar model was derived on the basis of the discrete equations of motion for the particles of the unit cell by using the single vector function $\{u(x,y,t), v(x,y,t), \varphi(x,y,t)\}$. The two-field model is derived by considering as a basis a macrocell consisting of two elementary cells. Although, all elements of the system are identical, they are marked with different numbers as shown in Figure 1b. Also, the generalized displacements are denoted as $u_{k,m}^{[n]}, v_{k,m}^{[n]}, \varphi_{k,m}^{[n]}$ where the superscript $n = 1, 2$ identifies the generalized displacements of the elements included in the macrocell. Two vector functions $\{u^{[n]}(x,y,t),\ v^{[n]}(x,y,t),\ \varphi^{[n]}(x,y,t)\}$, $n=1,2$, are used in the two-field theory to describe the displacements and rotations of the particles marked by numbers $n=1,2$, respectively. The behavior of the macrocell is governed by six discrete equations of motion. The application of Taylor series expansions leads to six continuum equations of the multifield model.

The following new field functions are introduced:

$$u^{[n]}(x,y,t)=u(x,y,t)+(-1)^n \tilde{u}(x,y,t),$$
$$v^{[n]}(x,y,t)=v(x,y,t)+(-1)^n \tilde{v}(x,y,t), \quad (12)$$
$$\varphi^{[n]}(x,y,t)=\varphi(x,y,t)+(-1)^n \tilde{\varphi}(x,y,t),\ n=1,2,$$

in order to decouple the six equations of the two-field model onto two subsystems.

The equations for the vector function $\{u(x,y,t),\ v(x,y,t),\ \varphi(x,y,t)\}$ coincide with Equations (9)-(11). The other three equations for $\{\tilde{u}(x,y,t),\ \tilde{v}(x,y,t),\ \tilde{\varphi}(x,y,t)\}$ have the following form

$$M\tilde{u}_{tt} = \left(-K_n + K_n^d\right)h^2\tilde{u}_{xx} + \left(-K_s + K_n^d\right)h^2\tilde{u}_{yy} + 2K_n^d h^2 \tilde{v}_{xy} - 4(K_n+K_s)\tilde{u}$$
$$- K_s h^2 \tilde{\varphi}_y - \frac{1}{12}(K_n - K_n^d)h^4 \tilde{u}_{xxxx} - \frac{1}{12}(K_s - K_n^d)h^4 \tilde{u}_{yyyy} + \frac{1}{2}K_n^d h^4 \tilde{u}_{xxyy} \quad (13)$$
$$+ \frac{1}{3}K_n^d h^4(\tilde{v}_{xxxy} + \tilde{v}_{xyyy}) - \frac{1}{6}K_s h^4 \tilde{\varphi}_{yyy},$$

$$M\tilde{v}_{tt} = 2K_n^d h^2 \tilde{u}_{xy} + \left(-K_s + K_n^d\right)h^2 \tilde{v}_{xx} + \left(-K_n + K_n^d\right)h^2 \tilde{v}_{yy} - 4(K_n+K_s)\tilde{v}$$
$$+ K_s h^2 \tilde{\varphi}_x - \frac{1}{12}(K_s - K_n^d)h^4 \tilde{v}_{xxxx} - \frac{1}{12}(K_n - K_n^d)h^4 \tilde{v}_{yyyy} + \frac{1}{2}K_n^d h^4 \tilde{v}_{xxyy} \quad (14)$$
$$+ \frac{1}{3}K_n^d h^4(\tilde{u}_{xxxy} + \tilde{u}_{xyyy}) + \frac{1}{6}K_s h^4 \tilde{\varphi}_{xxx},$$

$$I\tilde{\varphi}_{tt} = -\left(G_r - \frac{1}{4}h^2 K_s\right)h^2 (\tilde{\varphi}_{xx} + \tilde{\varphi}_{yy}) + K_s h^2 (\tilde{u}_y - \tilde{v}_x) - 8G_r \tilde{\varphi}$$
$$- \frac{1}{12}\left(G_r - \frac{1}{4}K_s h^2\right)h^4 (\tilde{\varphi}_{xxxx} + \tilde{\varphi}_{yyyy}) + \frac{1}{6}K_s h^4 (\tilde{u}_{yyy} - \tilde{v}_{xxx}). \quad (15)$$

Thus, the two-field model consists of equations (9)-(11) of the single-field theory and equations (13)-(15), the meaning of which will be clarified by the following analysis.

### 5. Plane wave solutions: the comparative analysis of the models

We are looking for solutions of the discrete equations of motion (3)-(5) in the following form

$$u_{k,m} = \tilde{U} \exp[i(\omega t - kK_x - mK_y)], \quad (16)$$

$$v_{k,m} = \widetilde{V} \exp[i(\omega t - kK_x - mK_y)], \tag{17}$$

$$\varphi_{k,m} = \widetilde{\Phi} \exp[i(\omega t - kK_x - mK_y)], \tag{18}$$

where $K_x = k_x h$, $K_y = k_y h$; $k_x$ and $k_y$ are wave numbers; $\widetilde{U}$, $\widetilde{V}$, and $\widetilde{\Phi}$ are amplitudes, and $\omega$ is the angular frequency.

By substituting expressions (16)-(18) into Equations (3)-(5) we obtain a system of three linear equations

$$(a_{11} + M\omega^2)\widetilde{U} + a_{12}\widetilde{V} + a_{13}\,i\widetilde{\Phi} = 0, \tag{19}$$

$$a_{12}\widetilde{U} + (a_{22} + M\omega^2)\widetilde{V} + a_{23}\,i\widetilde{\Phi} = 0, \tag{20}$$

$$a_{13}\widetilde{U} + a_{23}\widetilde{V} + (a_{33} + I\omega^2)\,i\widetilde{\Phi} = 0, \tag{21}$$

with coefficients

$$\begin{aligned}
a_{11} &= 2(\cos K_x - 1)K_n + 2(\cos K_y - 1)K_s + 2(\cos K_x \cos K_y - 1)K_n^d, \\
a_{12} &= -2K_n^d \sin K_x \sin K_y, \quad a_{13} = -h K_s \sin K_y, \quad a_{23} = hK_s \sin K_x, \\
a_{22} &= 2(\cos K_y - 1)K_n + 2(\cos K_x - 1)K_s + 2(\cos K_x \cos K_y - 1)K_n^d, \\
a_{33} &= -(2 + \cos K_x + \cos K_y)h^2 K_s / 2 + 2(\cos K_x + \cos K_y - 2)G_r,
\end{aligned} \tag{22}$$

In the continuum models, the following plane-wave solution is considered

$$u(x,y,t) = \widetilde{U}\exp[i(\omega t - k_x x - k_y y)], \tag{23}$$

$$v(x,y,t) = \widetilde{V}\exp[i(\omega t - k_x x - k_y y)], \tag{24}$$

$$\varphi(x,y,t) = \widetilde{\Phi}\exp[i(\omega t - k_x x - k_y y)], \tag{25}$$

which is the continuum analog of the discrete solution (16)-(18).

Substituting (23)-(25) in Equations (9)-(11) leads to Equations (19)-(21) with coefficients

$$\begin{aligned}
c_{11} &= (-K_x^2 + K_x^4/12)K_n + (-K_y^2 + K_y^4/12)K_s \\
&\quad + (-K_x^2 - K_y^2 + K_x^4/12 + K_x^2 K_y^2/2 + K_y^4/12)K_n^d, \\
c_{12} &= (-2K_x K_y + K_x K_y^3/3 + K_x^3 K_y/3)K_n^d, \\
c_{13} &= -(K_y - K_y^3/6)hK_s, \quad c_{23} = -c_{13}(x \leftrightarrow y), \quad c_{22} = c_{11}(x \leftrightarrow y), \\
c_{33} &= (-8 + K_x^2 + K_y^2 - K_x^4/12 - K_y^4/12)h^2 K_s / 4 \\
&\quad + (-K_x^2 - K_y^2 + K_x^4/12 + K_y^4/12)G_r.
\end{aligned} \tag{26}$$

These coefficients essentially can be recognized as the Taylor series expansions of the coefficients (22) of the discrete system around the point $(K_x, K_y) = (0, 0)$ up to the fourth order. In the case of classical micropolar model, the coefficients (26) include the terms of the Taylor series expansion of the coefficients in Equation (22) up to the second order. Accordingly, the dispersion curves of the conventional and higher-order gradient single-field models coincide with the dispersion curves of the discrete system at the point $(K_x, K_y) = (0, 0)$ and approximate them around this point. The higher-order gradient model improves the accuracy of the approximation at this point in comparison with the classical micropolar model. However, for short wavelength waves both single-field micropolar models produce results with an essential error.

The two-field model includes the equations of the single-field model, Equations (9)-(11), and additional equations (13)-(15). Six dispersion surfaces for the two-field model are defined in the area $|K_x \pm K_y| \leq \pi$. They can be split into two groups.

The surfaces of the first group correspond to those of the single-field model defined in the area $|K_x \pm K_y| \leq \pi$. Thus, two-field models possess the properties of single-field models and provide a good approximation of the dispersion surfaces for the discrete system for long wavelength waves.

The dispersion relations of the second group of surfaces correspond to Equations (13)-(15) of the two-field model. The substitution of expressions (23)-(25) into these equations leads to a system of Equations (19)-(21) with coefficients

$$\tilde{c}_{11} = (-4 + K_x^2 - K_x^4/12)K_n + (-4 + K_y^2 - K_x^4/12)K_s$$
$$+ (-K_x^2 - K_y^2 + K_x^4/12 + K_x^2 K_y^2/2 + K_y^4/12)K_n^d,$$
$$\tilde{c}_{12} = (-6K_x K_y + K_x^3 K_y + K_x K_y^3)K_n^d/3, \qquad (27)$$
$$\tilde{c}_{13} = (K_y - K_y^3/6)hK_s, \; \tilde{c}_{23} = -\tilde{c}_{13}(x \leftrightarrow y), \; \tilde{c}_{22} = \tilde{c}_{11}(x \leftrightarrow y),$$
$$\tilde{c}_{33} = (-K_x^2 - K_y^2 + K_x^4/12 + K_y^4/12)h^2 K_s/4$$
$$+ (-8 + K_x^2 + K_y^2 - K_x^4/12 - K_y^4/12)G_r.$$

These coefficients can be obtained by replacing $K_x \to -\pi + K_x$ and $K_y \to -\pi + K_y$ in the coefficients in Equation (22) for the discrete system and their Taylor series expansions around the point $(K_x, K_y) = (0, 0)$ up to fourth order terms. Because of invariance of the dispersion relations under the transformation of variables $(K_x, K_y) \leftrightarrow (-K_x, -K_y)$, the dispersion surfaces of the two-field model plotted for the coefficients (27) in the area $K_x + K_y \leq \pi$, $K_x \geq 0$, $K_y \geq 0$ being reflected about the line $K_x = \pi/2$, $K_y = \pi/2$ coincide with the dispersion surfaces of the discrete system at $(K_x, K_y) = (\pi, \pi)$ and approximate them in the area $K_x + K_y \geq \pi$, $K_x \leq \pi$, $K_y \leq \pi$ around this point. The inclusion of higher-order derivatives into the model improves the accuracy.

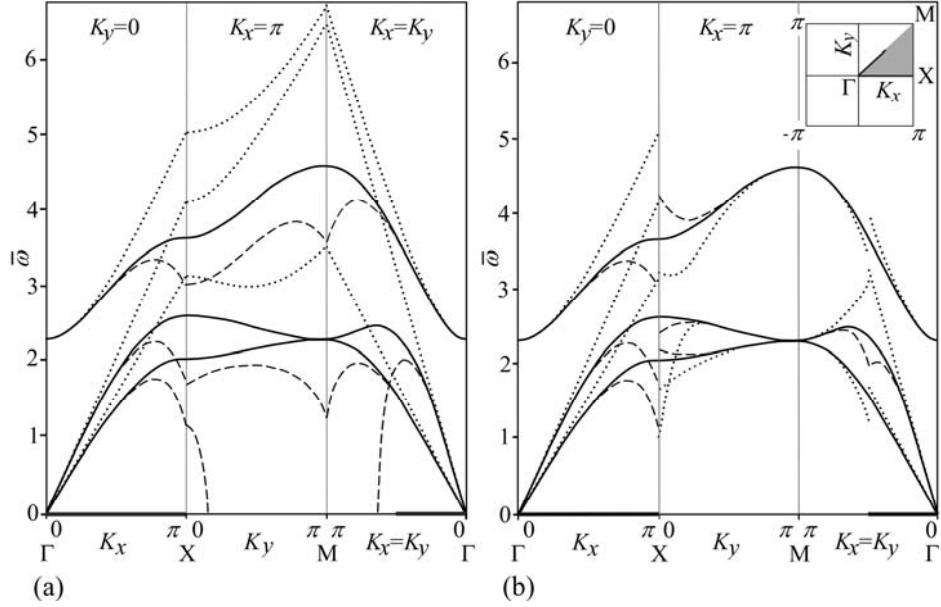

**Figure 2.** The dispersion curves in the sections $K_y = 0$, $K_x = \pi$, $K_x = K_y$ of the dispersion surfaces for discrete system (solid lines). The same curves obtained by using the single-field (a) and two-field (b) models with derivatives up to second (dotted lines) and fourth orders (dashed lines).

Figure 2 illustrates the results of the comparative analysis of the models, the accuracy of the approximation of the dispersion surfaces of the discrete system through the dispersion surfaces of single-field (Figure 2a) and two-field (Figure 2b) models, and the influence on accuracy of the order of derivatives in the models. The dispersion curves for the discrete system in the sections $K_y = 0$, $K_x = \pi$, $K_x = K_y$ are represented by solid lines. The same curves obtained for the field models with derivatives up to second and fourth orders are represented by dotted and dashed lines, respectively. We fix the parameters $K_n$, $h$, $M$ and consider dimensionless quantities $\overline{K}_s = K_s / K_n = 1/3$, $\overline{K}_n^d = K_n^d / K_n = 0.74$, $\overline{G}_r = G_r / K_n h^2 = 1/3$, $\overline{I} = I / M h^2 = 1/8$. Frequency is also presented in dimensionless form according to the following expression $\overline{\omega} = \omega \sqrt{M / K_n}$. The curves for the single-field and the two-field models coincide in the area $K_x + K_y \leq \pi$ and give a good approximation of the curves for the discrete system in the area of the long wavelength waves with wave numbers around $\Gamma$-point. The two-field models give additionally good approximation for short waves in area around M-point, where the single-field models give significant inaccuracy. The models containing the derivatives up to fourth order improve the approximation in comparison with the models with derivatives up to the second order.

### 6. One-dimensional models

We consider the one-dimensional deformations of a lattice placed between two rigid components (see Figure 3). Assuming that the generalized displacements are constant for elements along the diagonals, that is, for $k+m=const$, we denote components $U_{k,m}$, $V_{k,m}$, and $\Phi_{k,m}$ by using the abbreviated notations $U_m$, $V_m$, and $\Phi_m$. The equations for $U_m$ and $V_m$, $\Phi_m$ are decoupled, and we will concentrate on the solutions for $U_m$ only.

By introducing the new coordinates $O\xi\eta$ and performing a change of variables in Equations (3)-(5), one obtains the discrete equation of motion for one-dimensional deformations

$$M\ddot{U}_m = (K_n + K_s)(U_{m-1} - 2U_m + U_{m+1}) + K_n^d(U_{m-2} - 2U_m + U_{m+2}). \quad (28)$$

By changing the variables $(x,y) \to (\xi,\eta)$ in Equations (9)-(11) and considering one-dimensional displacements, we obtain the single-field higher-order gradient variant of Equation (28)

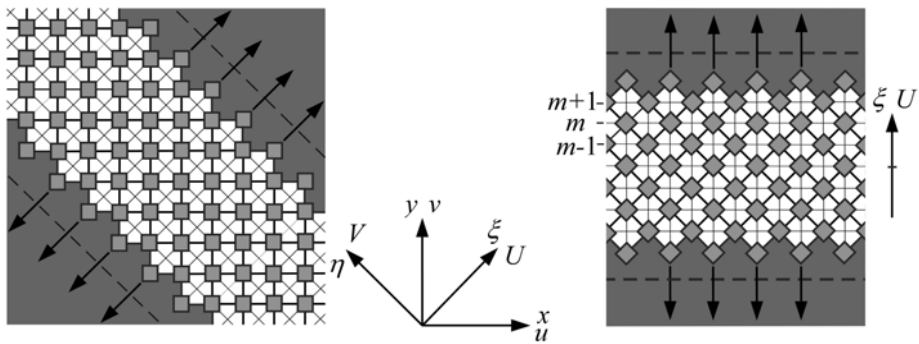

**Figure 3.** Representation of a thin lattice layer in a problem of tension between two rigid parts in different coordinate systems.

$$MU_{tt} = (K_n + K_s + 4K_n^d)H^2 U_{\xi\xi} + \frac{1}{12}(K_n + K_s + 16K_n^d)H^4 U_{\xi\xi\xi\xi}, \tag{29}$$

where $H = \sqrt{2}h/2$ is distance between layers. Equation (29) can be obtained independently by using Taylor series expansions in Equation (28).

Similarly, changing the variables $(x,y) \to (\xi,\eta)$ in Equations (13)-(15) and considering one-dimensional displacements leads to the second equation of the two-field model

$$M\widetilde{U}_{tt} = -4(K_n + K_s)\widetilde{U} - (K_n + K_s - 4K_n^d)H^2 \widetilde{U}_{\xi\xi} - \frac{1}{12}(K_n + K_s - 16K_n^d)H^4 \widetilde{U}_{\xi\xi\xi\xi}. \tag{30}$$

In order to explain the notations and to illustrate the method of derivation of the multifield models, we will obtain Equations (29) and (30) of the two-field model directly from Equation (28).

Although the unit cell in the problem under consideration consists of the single layer, we assume that a cell of periodicity consists of two layers and use the notations $U_{2n}^{[1]}(t)$ and $U_{2n+1}^{[2]}(t)$ with different superscripts [1] and [2] for displacements of the layers with coordinates $\xi = 2nH$ and $\xi = (2n+1)H$, respectively. Equation (28) can thus be rewritten in the form

$$M\ddot{U}_{2n}^{[1]} = (K_n + K_s)(U_{2n-1}^{[2]} - 2U_{2n}^{[1]} + U_{2n+1}^{[2]}) + K_n^d(U_{2n-2}^{[1]} - 2U_{2n}^{[1]} + U_{2n+2}^{[1]}), \tag{31}$$

$$M\ddot{U}_{2n+1}^{[2]} = (K_n + K_s)(U_{2n}^{[1]} - 2U_{2n+1}^{[2]} + U_{2n+2}^{[1]}) + K_n^d(U_{2n-1}^{[2]} - 2U_{2n+1}^{[2]} + U_{2n+3}^{[2]}). \tag{32}$$

We use two functions $U^{[1]}(x,t)$ and $U^{[2]}(x,t)$ in order to describe displacements of odd and even layers

$$U^{[1]}(\xi,t)\big|_{\xi=2nH} = U_{2n}^{[1]}(t), \quad U^{[2]}(\xi,t)\big|_{\xi=(2n+1)H} = U_{2n+1}^{[2]}(t), \tag{33}$$

The Taylor series expansions of the displacements in Equations (31) and (32) up to fourth order terms around the points for which these equations were obtained, gives the system of coupled equations for the two-field model

$$MU_{tt}^{[1]} = \overline{L}U^{[1]} - \widetilde{L}(U^{[1]} - U^{[2]}), \tag{34}$$

$$MU_{tt}^{[2]} = \overline{L}U^{[2]} + \widetilde{L}(U^{[1]} - U^{[2]}), \tag{35}$$

where we separate the operator for the single-field model, Equation (29),

$$\overline{L} = (K_n + K_s + 4K_n^d)H^2 \frac{\partial^2}{\partial \xi^2} + \frac{1}{12}(K_n + K_s + 16K_n^d)H^4 \frac{\partial^4}{\partial \xi^4} \tag{36}$$

from the additional operator

$$\widetilde{L} = (K_n + K_s)\left(2 + H^2 \frac{\partial^2}{\partial \xi^2} + \frac{1}{12} H^4 \frac{\partial^4}{\partial \xi^4}\right), \tag{37}$$

which describes the interaction of the fields. This representation of the model can be useful for the interpretation and the generalization of two-field models, in particular in the presence of nonlinearities. In the linear case, it is convenient to split the system of coupled equations (34) and (35) in two independent equations (29) and (30) by introducing the new field functions

$$U = \frac{1}{2}\left(U^{[2]} + U^{[1]}\right), \quad \widetilde{U} = \frac{1}{2}\left(U^{[2]} - U^{[1]}\right). \tag{38}$$

### 7. One-dimensional dynamic problem: comparative analysis of models

**7.1. *Discrete model.*** We consider solutions of the discrete equation of motion (28) of the form

$$U_m(t) = \widetilde{U} e^{i\omega t - Km} \tag{39}$$

with complex values $K = K_{Re} + iK_{Im}$. Substituting Equation (39) into (28) leads to the relation

$$\overline{\omega}^2 = 2(1 - \cosh K) + 2\gamma(1 - \cosh 2K), \tag{40}$$

where $\overline{\omega} = \omega\sqrt{M/(K_n + K_s)}$ and $\gamma = K_n^d/(K_n + K_s)$.

Substituting $K = iK_{Im}$ in Equation (40), and letting $\Omega = \overline{\omega}^2$, $Z = 4\sin^2(K_{Im}/2)$ gives

$$\Omega = (1 + 4\gamma)Z - \gamma Z^2. \tag{41}$$

The analog (41) of the dispersion relation (40) is useful for the analysis of the dispersion curves of the discrete system because there is a remarkable correspondence between the curves $\overline{\omega} = \overline{\omega}(K_{Im}, K_{Re})$ in three-dimensional space $\omega \geq 0$, $0 \leq K_{Im} \leq \pi$, $K_{Re} \geq 0$ and the curve $\Omega = \Omega(Z)$ in the two-dimensional space $\Omega \geq 0$, $-\infty < Z < \infty$. Namely, the dispersion curves $\omega = \omega(K_{Im}, K_{Re})$ in the planes $K_{Im} = 0$, $K_{Re} = 0$, and $K_{Im} = \pi$ correspond to parts of the parabola $\Omega = \Omega(Z)$ in the intervals $Z < 0$, $0 < Z < 4$, and $Z > 4$, respectively. The branch $\omega = \omega(K_{Im}, K_{Re})$ defined in the area of complex values ($K_{Im} \neq 0$, $K_{Re} \neq 0$) is located at frequencies $\Omega$, at which there are no points of the curve $\Omega = \Omega(Z)$.

The complex dispersion relation (40) determines three dispersion curves of qualitatively different solutions.

The curve in the plane $K_{Re} = 0$ is defined by the relation

$$\overline{\omega}^2 = 2(1 - \cos K_{Im}) + 2\gamma(1 - \cos 2K_{Im}). \tag{42}$$

It corresponds to points of the parabola inside the interval $0 < Z < 4$. This is the branch corresponding to harmonic solutions $\overline{\omega} = \overline{\omega}(K_{Im}, 0)$.

The curve in the plane $K_{Im} = \pi$

$$\overline{\omega}^2 = 2(1 + \cosh K_{Re}) + 2\gamma(1 - \cosh 2K_{Re}) \tag{43}$$

corresponds to points of parabola for the values $Z > 4$. This is the branch associated with rapidly varying spatially localized solutions, which correspond to evanescent waves.

The branch $\overline{\omega} = \overline{\omega}(K_{Im}, K_{Re})$ for the complex values $K_{Re} \neq 0$, $K_{Im} \neq 0$ is defined in the parametric form $K_{Re} = K_{Re}(\overline{\omega})$, $K_{Im} = K_{Im}(\overline{\omega})$ by the equations

$$\cos K_{Im} \cosh K_{Re} = -1/4\gamma,$$
$$\cos^2 K_{Im} + \cosh^2 K_{Re} = (\overline{\omega}^2 - 2)/4\gamma. \tag{44}$$

It is located in the area of frequencies $\Omega$ above the value $\overline{\Omega}_{max} = (2\sqrt{\gamma} + 1/2\sqrt{\gamma})^2$. This value corresponds to the maximum of the parabola, located at $Z_{max} = 2 + 1/2\gamma$, which belong to the interval $2 < Z_{max} < 4$ for $\gamma > 1/4$ or to the half-line $Z_{max} > 4$ for $\gamma < 1/4$. Hence, the curve (44) begins at the point of maximum of the curve defined by Equation (42) for $\gamma > 1/4$ or Equation (43) for $\gamma < 1/4$.

Since the parabola is not defined for $Z < 0$, $\Omega > 0$, there are no dispersion curves of the discrete system in the plane $K_{Im} = 0$, and, accordingly, Equation (28) does not possess slowly varying spatially localized solutions of the form (39).

**7.2. Single-field models.** The analysis of the single- and two-field models is based on the following solution

$$U(\xi, t) = \overline{U} e^{i\omega t - K\xi/H}, \tag{45}$$

which is the analog of the discrete solution (39) in the continuum case.

The substitution of expression (45) into Equation (29) of the single-field model with derivatives up to the second order leads to the dispersion relation for harmonic solutions

$$\overline{\omega}^2 = (1 + 4\gamma)K_{Im}^2. \tag{46}$$

The corresponding curve belongs to the plane $K_{Re} = 0$ and defines the tangent line to the dispersion curve of the discrete system defined by Equation (42) at the point $\omega = 0$, $K_{\text{Im}} = 0$. For short wavelength waves in the area $K_{\text{Im}} \approx \pi$ the model shows considerable inaccuracy.

The single-field model with derivatives up to fourth order, Equation (29), gives the following dispersion relation in the case $K_{\text{Re}} = 0$

$$\overline{\omega}^2 = (1+4\gamma)K_{\text{Im}}^2 - (1+16\gamma)K_{\text{Im}}^4/12, \qquad (47)$$

which corresponds to a Taylor series expansion up to the fourth order of the dispersion relation of the discrete system (42) around $K_{\text{Im}} = 0$. From a physical point of view, this means that the higher-order gradient model describes the dispersive behavior. This model has better accuracy over a larger range of wavelengths with respect to the model with derivatives up to the second order, but still suffers from severe inaccuracies in the short wavelength limit.

The single-field model with derivatives up to the fourth order, Equation (29), characterized by dispersion curve in the complex plane ($K_{\text{Im}} \neq 0$, $K_{\text{Re}} \neq 0$) similar to the dispersion curve of the discrete system defined by Equation (44). However, inaccuracy in the description of this curve is large because it begins at the point of maximum of the curve, defined by Equation (42) or Equation (43). This point belongs to the area of middle and short waves. As previously discussed, the single-field model for these waves is affected by considerable inaccuracy.

It should be noted that both single-field models do not provide dispersion curves in the plane $K_{\text{Im}} = \pi$ which are similar to those of the discrete system defined by Equation (43). Hence, single-field models do not capture the spatially localized short wavelength solutions.

**7.3. Two-field models.** The two-field model, Equations (29) and (30), includes the equation of the single-field model, Equation (29). Therefore, like the single-field model, the two-field model is able to capture the properties of the discrete system in the range $0 \leq K_{\text{Im}} < \pi/2$, where the single-field model has a good accuracy.

Equations (30) of the two-field model in the case when only derivatives up to the second order are taken into account leads to the following relation

$$\overline{\omega}^2 = 4 + (1-4\gamma)K^2. \qquad (48)$$

This complex dispersion relation gives two branches

$$\overline{\omega}^2 = 4 - (1-4\gamma)K_{\text{Im}}^2, \qquad (49)$$

$$\overline{\omega}^2 = 4 + (1-4\gamma)K_{\text{Re}}^2 \qquad (50)$$

in the area $0 \leq K_{\text{Im}} < \pi/2$, $K_{\text{Re}} \geq 0$.

The relation (49) defines dispersion curve in the plane $K_{Re} = 0$. It can be obtained by replacing $K_{Im} \to \pi - K_{Im}$ in Equation (42) and its Taylor series expansion around the point $K_{Im} = 0$ up to second order terms. Hence, the dispersion curve defined by Equation (49), reflected about the line $K_{Im} = \pi/2$, approximates the branch of harmonic wave solutions of the discrete system, Equation (42), in the area of short wavelength around the point $K_{Im} = \pi$.

The relation (50) defines the dispersion curve in the plane $K_{Im} = 0$, which, after reflection about the plane $K_{Im} = \pi/2$, approximates the curve for short wavelength localized solutions of the discrete system, Equation (43), in the plane $K_{Im} = \pi$ around the point $K_{Re} = 0$.

By taking into consideration derivatives up to the fourth order in the additional Equation (30) for the two-field model, we can improve approximation for the branches of the harmonic, Equation (42), and spatially localized, Equation (43), short wavelength solutions of the discrete system at $(K_{Im}, K_{Re}) = (\pi, 0)$ up to the fourth order terms. The corresponding approximate relations have the form

$$\overline{\omega}^2 = 4 - (1 - 4\gamma)(K_{Im} - \pi)^2 + (1 - 16\gamma)(K_{Im} - \pi)^4 / 12, \tag{51}$$

$$\overline{\omega}^2 = 4 + (1 - 4\gamma)K_{Re}^2 + (1 - 16\gamma)K_{Re}^4 / 12. \tag{52}$$

Thus, the two-field model has the properties of the single-field model in the area of long waves ($K_{Im} \approx 0$) and additionally demonstrates a good accuracy as for the short wavelength harmonic wave solutions ($K_{Im} \approx \pi$, $K_{Re} = 0$) and for short wavelength solutions with weak spatial localization ($K_{Im} = \pi$, $K_{Re} \approx 0$). Taking into account derivatives up to the fourth order in the equations of the two-field model improves the accuracy around $(K_{Im}, K_{Re}) = (0, 0)$ and $(K_{Im}, K_{Re}) = (\pi, 0)$. For $\gamma > 1/16$ the two-field model with derivatives up to the fourth order gives also the dispersion curve in the area of complex values, $K_{Im} \neq 0$, $K_{Re} \neq 0$, similar to the curve defined by Equation (44) for the discrete system.

Spatially localized short-wavelength, high frequency excitations, which exist and propagate in nonlinear discrete systems, are known as intrinsic localized modes, or discrete breathers [Sievers and Takeno 1988; Flach and Willis 1998]. The development of continuum models capable of providing good descriptions for both long and short wavelength phenomena gives an opportunity to study them separately and to investigate their potential interactions.

Figure 4 illustrates the results of the analysis. The dispersion curves in the planes of the harmonic $(K_{Re} = 0)$ and spatially localized short wavelength $(K_{Im} \approx \pi)$ solutions for the discrete system (solid lines), the two-field models with derivatives up to the second (dotted lines) and fourth (dashed lines) orders are presented for $\gamma = 1/2$. In the area of long wavelength solutions $(K_{Im} < \pi/2,\ K_{Re} = 0)$, the dispersion curves of the two-field model coincide with the dispersion curves of the single-field model and give a good approximation of the dispersion curves of the discrete system in the area of long waves $(K_{Im} \approx 0)$. Moreover, the two-field model produces branches that approximate the dispersion curves of the discrete system corresponding to the short wavelength harmonic and spatially localized solutions around the point $(K_{Im}, K_{Re}) = (\pi, 0)$. The points of intersection of the dispersion curses with the axis $\bar{\omega} = 0$ for different models are depicted by small circles. These points correspond to static solutions, which are analyzed in the next Section.

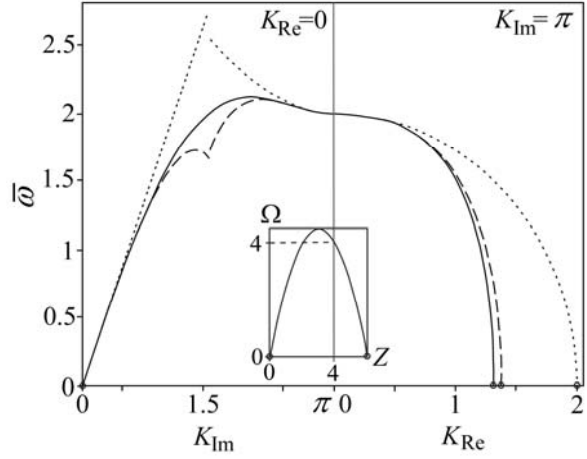

**Figure 4.** Dispersion curves of discrete system for harmonic, $K_{Re} = 0$, and localized short-wave solutions, $K_{Im} = \pi$, are shown by solid lines. The same curves obtained by using two-field models with derivatives up to second and fourth orders are presented by dotted and dashed lines, respectively.

## 8. Static one-dimensional solutions

**8.1. *Discrete model.*** The parabola defined by Equation (41) intersects the axis $\Omega = 0$ at $Z_1 = 0$ and $Z_2 = 4 + 1/\gamma$. Since $Z_2 > 4$, the point where the dispersion curve intersects the plane $\bar{\omega} = 0$ belongs to the line $K_{Im} = \pi$ and, hence, the corresponding static solution has rapidly varying exponentially localized form.

The characteristic equation of the discrete equation (28) in the static problem has roots

$$K_{\text{Im}} = 0, \ K_{\text{Re}} = 0 \tag{53}$$

and

$$K_{\text{Im}} = \pi, \ K_{\text{Re}} = \pm \lambda, \tag{54}$$

where

$$\lambda = \ln\left[(1+1/2\gamma) + \sqrt{(1+1/2\gamma)^2 - 1}\right] \tag{55}$$

is the solution of the equation

$$1 + \cosh\lambda + \gamma(1 - \cosh 2\lambda) = 0. \tag{56}$$

The general static solution of Equation (28) has the form

$$U_m = C_0 + mC_1 + (-1)^m e^{\lambda m} C_2 + (-1)^m e^{-\lambda m} C_3. \tag{57}$$

This solution consists of linear and rapidly varying terms that are defined by the roots (53) and (54), respectively. The parameter $\lambda$ defines the degree of localization of the solution, and the constants $C_n$ are defined by the particular set of assigned boundary conditions.

**8.2.** *Single-field models.* The results of the comparative analysis between discrete and field models for dynamic solutions (Section 7) will be used here for comparison of the static solutions.

Single-field models with derivatives up to second and fourth orders both approximate the branch of the discrete system defined by Equation (42) for long waves, $K_{\text{Re}} = 0$, $K_{\text{Im}} \approx 0$, at low frequencies, $\bar{\omega} \approx 0$. Therefore, the characteristic polynomial for the static equations of the single-field models has trivial solution $K_{\text{Re}} = 0$, $K_{\text{Im}} = 0$, Equation (53), of the second order. Consequently, both single-field models give the linear slowly varying part of the solution (57).

Due to the fact that the equation (29) in case when derivatives up to fourth order are taken into account gives a polynomial of the fourth order for static solutions, additional roots are found in the plane $\bar{\omega} = 0$. Because of that, one may expect that the single field model in this case gives the rapidly varying exponentially localized part of the static solution of the discrete system. However, as it was established during the previous analysis on the dynamic solutions, Section 7.2, the single-field models do not predict the branch of spatially localized solutions (43) with root (54). The additional roots of the model belong to the branch of the harmonic solutions in the plane $K_{\text{Re}} = 0$.

Thus, single-field models allow us to find only the linear slowly varying part of the static solution, but they do not predict the rapidly varying spatially localized solutions.

**8.3.** *Two-field model.* The continuum approximation for the static solution for the discrete system (57) can be found through Equations (29) and (30) obtained for the two-field model with derivatives up to the second order.

Equation (29) gives the following linear static solution

$$U(\xi) = c_0 + c_1 \xi / H . \tag{58}$$

By using Equation (30) it is possible to find the localized part of the solution

$$\widetilde{U}(\xi) = e^{\Lambda \xi / H} c_2 + e^{-\Lambda \xi / H} c_3 , \tag{59}$$

where $\Lambda$ is found from Equation (50) in the case $\overline{\omega} = 0$, that is,

$$4 + (1 - 4\gamma)\Lambda^2 = 0 . \tag{60}$$

This equation gives the point $K_{\mathrm{Re}}$ of the intersection of the plane $\overline{\omega} = 0$ by the dispersion curve of the two-field model defined by Equation (50) in the plane $K_{\mathrm{Im}} = \pi$.

From Equation (38)

$$U^{[n]}(\xi) = U(\xi) + (-1)^n \widetilde{U}(\xi), \tag{61}$$

we obtain a static solution

$$U^{[n]}(\xi) = c_0 + c_1 \xi / H + (-1)^n e^{\Lambda \xi / H} c_2 + (-1)^n e^{-\Lambda \xi / H} c_3 , \tag{62}$$

where $n = 1, 2$.

Thus, the continuum solution (62), obtained by using the two-field model is qualitatively similar to the discrete solution (57). It contains the linear part as well as the localized one with rapidly varying envelope. In regards to the quantitative comparison, let us note that the parameters $\lambda$ in the discrete and $\Lambda$ in the continuum solutions are equal to the values $K_{\mathrm{Re}}$ of intersections of the plane $\omega = 0$ with the dispersion curves defined by Equations (43) and (50), respectively. As established in Section 7.3, the curve defined by Equation (50) of the two-field model approximates the curve defined by Equation (43) of the discrete model at the point $(K_{\mathrm{Im}}, K_{\mathrm{Re}}) = (\pi, 0)$. Furthermore, Equation (60) for the parameter $\Lambda$ can be obtained from Equation (56) for $\lambda$ by using its Taylor series expansion up to the second order terms

$$2(1 + \cosh \lambda) + 2\gamma(1 - \cosh 2\lambda) = 4 + (1 - 4\gamma)\lambda^2 + O(\lambda^4). \tag{63}$$

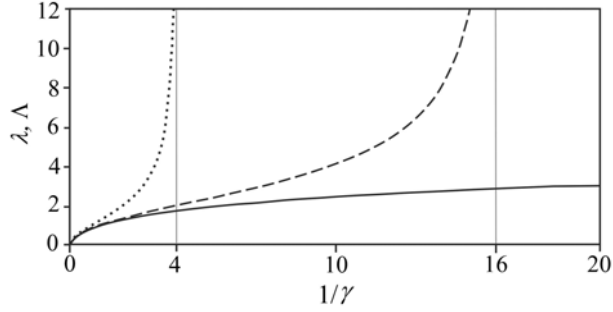

**Figure 5.** The dependencies of the localization parameters, $\lambda$ and $\Lambda$, on the parameter $1/\gamma = (K_n + K_s)/K_n^d$ of the discrete system calculated by using the discrete model (solid line) and two-field models with derivatives up to second (dotted line) and fourth (dashed line) orders.

Thus, we have demonstrated that the short wavelength spatially localized static deformations, which could not be obtained through the single-field approach (see Section 8.2), can be found by using the two-field model.

The accuracy of the two-field model for the derivation of static solutions is good in the case of weak localization. The solution (62) for the model in the case when only derivatives up to the second order are taken into account exists for $1/\gamma < 4$. While for $1/\gamma > 4$ the solution of the discrete system is highly localized. This explains the lack of accuracy of the two-field model, when the parameter $1/\gamma$ approaches its threshold value $1/\gamma \to 4$, and the reason why the solutions cannot be found when $1/\gamma > 4$.

Figure 5 shows the dependence of the localization parameters $\lambda$ and $\Lambda$ on the parameter $1/\gamma = (K_n + K_s)/K_n^d$ of the discrete system, calculated from the discrete model (solid line) and from the two-field models with derivatives up to second (dotted line) and fourth (dashed line) orders. The inclusion of fourth order derivatives increases both the accuracy of the derivation of the localization parameter $\lambda$, and the region of parameters $\gamma$ where spatially localized static solution can be found. Let us note that there exist systems with small parameter $\lambda$. In such systems short wavelength solutions are weakly localized. The two-field model with derivatives up to the second order produces rather exact results and there is no need to use higher-order gradient two-field model in this case.

The static solution for tension (compression) of a layer between two rigid parts (Figure 3) with boundary conditions

$$U_{-N+1} = -U_*, \quad U_{-N+2} = -U_*, \quad U_{N-1} = U_*, \quad U_N = U_* \qquad (64)$$

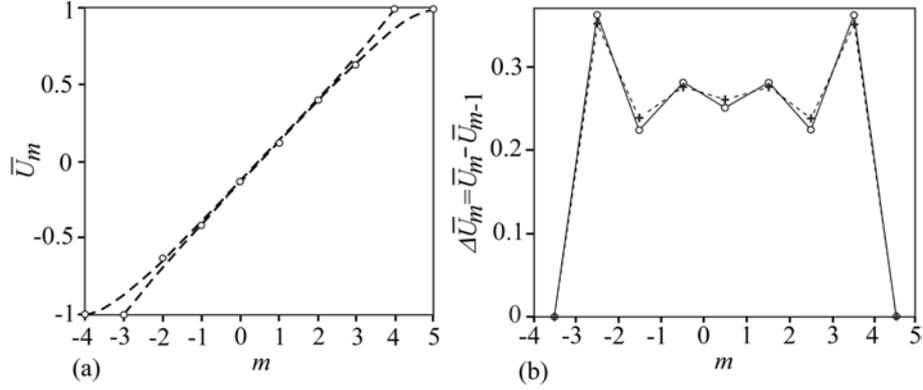

**Figure 6.** (a) Displacements of layers of the lattice are shown by circles. Their approximations by using two slowly varying functions in the two-field model are presented by dashed lines. (b) The differences of the displacements of neighboring elements calculated by using the discrete and two-field models are presented by circles and crosses, respectively. Continuous and dashed lines are drawn to underline short wavelength behavior of the solutions near the boundaries.

is presented in Figure 6a. We choose the dimensionless parameter $\gamma = 1$. The number of layers is equal to ten, that is, $N = 5$. The displacements are given in dimensionless form $\overline{U}_m = U_m / U_*$. The displacements of the discrete system are found by using the static solution (57) of the discrete equation (28) with boundary conditions (64). They are shown by small circles in Figure 6a. The displacements obtained through the two-field solution (62) are represented as dashed lines. In order to underline the short-wavelength behaviour of the solutions near the boundaries, the difference of displacements $\Delta \overline{U}_m = \overline{U}_{m+1} - \overline{U}_m$ is shown in Figure 6b. The values calculated by using discrete and two-field models are represented as circles and crosses, respectively. Figure 6a demonstrates that both slowly varying displacements in the center and rapidly varying displacements near the boundaries in the discrete system are effectively described by two smooth field functions. The approximated models were derived through the Taylor series expansion (8) of the displacements $U(x \pm mH, t)$ keeping derivatives of the lowest order under assumption that $U(x,t)$ varies slowly with respect to spatial variable. This may help qualitatively understand why two-field models provide good approximations for both slowly and rapidly varying displacements. When we try to describe the short wavelengths deformations by using a single function it should rapidly vary in the corresponding areas. The sin-gle-field models with lower gradient terms do not capture the rapidly varying static localized solutions and may lead to significant errors in the prediction of displacements in areas where such solutions may take place (near boundaries, defects, localized forces, and so on). The two-field model makes it possible to describe these effects within the framework of the field theory.

We have considered the application of the two-field model, Equations (9)-(11) and (13)-(15), for the solution of one-dimensional problem (Figure 3) in order to demonstrate its advantages in comparison with single-field models. Another interesting problem is the determination of the shear deformations of the layer in the case of transverse displacements of the rigid parts in different directions. In this case there exists a range of parameters of the discrete system for which smooth exponential localization of deformations takes place. The deformations in this case can be studied by using the single-field theory. However, for other parameters, the localized deformations have short wavelength form and the equations of the single-field theory are not sufficient to capture them. The two-field model, Equations (9)-(11) and (13)-(15), enables us to study deformations in both cases. The analysis of this problem deserves a dedicated study, which is currently being carried out by the authors, and which will be the subject of a future paper.

## 9. Conclusions

The classical model for Cosserat media has wide applications in problems where not only displacements, but also rotations of structural elements should be taken into account. Rotational degrees of freedom naturally appear, for example, for bodies with elements having finite sizes and bodies with beam-like microstructure. Some examples are granular media [Limat 1988; Mühlhaus and Oka 1996; Pasternak and Mühlhaus 2000], beam lattices [Noor 1988], masonry walls [Casolo 2004], auxetics [Lakes 1991; Vasiliev et al. 2002; Grima et al. 2007], bodies with chiral structure [Spadoni and Ruzzene 2007], liquid crystals, dielectric crystals [Pouget et al. 1986; Askar 1986; Maugin 1999], thin films [Randow et al. 2006], among others.

The discrete model of square lattice of elements with rotational degrees of freedom is used to obtain generalized continuum models which describe essential structural effects in Cosserat media with microstructure. We utilize ideas and methods of the well-developed micropolar and higher-order gradient theories and of the multifield theory, which is still being investigated and expanded, and consider the possibilities and advantages of their applications separately and in combination. The derivation of the continuum models starting from structural model gives us the possibility to test their accuracy, and to study and compare their properties. We have shown that their application allows describing qualitatively different effects in bodies with microstructure. By increasing the number of fields, the multifield approach gives a natural way to describe both long- and short wavelength deformations. The latter ones are often considered as inacces-sible for continuum models. However, it should be noted that such deformations in some cases may become very important, in particular in fracture, instability, and plasticity problems. Their description gives a possibility to study similar phenomena in the framework of generalized continuum mechanics.

## Acknowledgements

The work of A.E.M. was supported by Australian Research Council.

## References

[Aifantis 2003] E.C. Aifantis "Update on a class of gradient theories" *Mechanics of Materials* **35** (2003), 259-280.

[Askar 1986] A. Askar, *Lattice dynamical foundations of continuum theories*, World Scientific, Singapore, 1986.

[Askes et al. 2002] H. Askes, A.S.J. Suiker, L.J. Sluys, "A classification of higher-order strain-gradient models – linear analysis", *Archive of Applied Mechanics* **72** (2002), 171-188.

[Bažant and Jirásek 2002] Z.P. Bažant, M. Jirásek, "Nonlocal integral formulation of plasticity and damage: survey of progress", *Journal of Engineering Mechanics* **128**:11 (2002), 1119-1149.

[Born and Huang 1954] M. Born, K. Huang, *Dynamical theory of crystal lattices*, Clarendon Press, Oxford, 1954.

[Casolo 2004] S. Casolo, "Modelling in-plane micro-structure of masonry walls by rigid elements", *International Journal of Solids and Structures* **41** (2004), 3626-3641.

[Cosserat and Cosserat 1909] E. Cosserat, F. Cosserat, *Théorie des Corps Déformables*, Hermann A. et Fils, Paris, 1909.

[Eringen 1999] A.C. Eringen, *Microcontinuum Field Theories: Foundations and Solids,* Springer-Verlag, New York, 1999.

[Flach and Willis 1998] S. Flach, C.R. Willis, "Discrete breathers", *Physics Reports* **295** (1998), 181-264.

[Fleck and Hutchinson 1997] N.A. Fleck, J.W. Hutchinson, "Strain gradient plasticity", *Advances in Applied Mechanics* **33** (1997), 295-361.

[Fleck and Hutchinson 2001] N.A. Fleck, J.W. Hutchinson, "A reformulation of strain gradient plasticity", *Journal of the Mechanics and Physics of Solids* **49** (2001), 2245-2271.

[Grima et al. 2007] J.N. Grima, V. Zammit, R. Gatt, A. Alderson, K.E. Evans, "Auxetic behaviour from rotating semi-rigid units", *Physica Status Solidi B* **244**:3 (2007), 866-882.

[Lakes 1991] R. Lakes, "Deformation mechanisms in negative Poisson's ratio materials: structural aspects", *Journal of Materials Science* **26** (1991), 2287-2292.

[Limat 1988] L. Limat, "Percolation and Cosserat elasticity: Exact results on deterministic fractal", *Physical Review B* **37** (1988), 672-675.

[Lomakin 1970] V.A. Lomakin, *Statistical problems of the mechanics of deformable solids* (in Russian), Nauka, Moscow, 1970.

[Maugin 1999] G.A. Maugin, *Nonlinear waves in elastic crystals*, Oxford Mathematical Monographs, OUP, 1999.

[Mühlhaus and Oka 1996] H.-B. Mühlhaus, F. Oka, "Dispersion and wave propagation in discrete and continuous models for granular materials", *International Journal of Solids and Structures* **33** (1996), 2841-2858.

[Noor 1988] A.K. Noor, "Continuum modelling for repetitive lattice structures", *Applied Mechanics Reviews* **41**:7 (1988), 285-296.

[Pasternak and Mühlhaus 2000] E. Pasternak, H.-B. Mühlhaus, "Cosserat and non-local continuum models for problems of wave propagation in fractured materials", in: Zhao, X.L., Grzebieta, R.H. (eds.), *Structural Failure and Plasticity - IMPLAST2000*, Pergamon, Amsterdam, pp. 741-746, 2000.

[Pavlov et al. 2006] I.S. Pavlov, A.I. Potapov, G.A. Maugin, "A 2D granular medium with rotating particles", *International Journal of Solids and Structures* **43** (2006), 6194-6207.

[Peerlings et al. 2001] R.H.J. Peerlings, M.G.D. Geers, R. de Borst, W.A.M. Brekelmans, "A critical comparison of nonlocal and gradient enhanced softening continua", *International Journal of Solids and Structures* **38** (2001), 7723-7746.

[Pouget et al. 1986] J. Pouget, A. Askar, G.A. Maugin, "Lattice model for elastic ferroelectric crystals: continuum approximation", *Physical Review B* **33** (1986), 6320-6325.

[Randow et al. 2006] C.L. Randow, G.L. Gray, F. Costanzo, "A direct continuum model of micro- and nano-scale thin films", *International Journal of Solids and Structures* **43** (2006), 1253-1275.

[Rogula 1985] D. Rogula, "Non-classical material continua", in: Niordson F.I., Olhoff N. (eds.), *Theoretical and Applied Mechanics, IUTAM,* Elsevier Science Publishers B.V., North Holland, 1985, pp. 339-353.

[Sievers and Takeno 1988.] A.J. Sievers, S. Takeno, "Intrinsic localized modes in anharmonic crystals", *Physical Review Letters* **61** (1988), 970-973.

[Spadoni and Ruzzene 2007] A. Spadoni, M. Ruzzene, "Numerical and experimental analysis of static compliance of chiral truss-core airfoils", *Journal of Mechanics of Materials and Structures* **2**:5 (2007), 965-981.

[Suiker et al. 2001] A.S.J. Suiker, A.V. Metrikine, R. de Borst, "Comparison of wave propagation characteristics of the Cosserat continuum model and corresponding discrete lattice models", *International Journal of Solids and Structures* **38** (2001), 1563-1583.

[Suiker and de Borst 2005] A.S.J. Suiker, R. de Borst, "Enhanced continua and discrete lattices for modelling granular assemblies", *Philosophical Transactions of the Royal Society A* **363** (2005), 2543-2580.

[Triantafyllidis and Bardenhagen 1993] N. Triantafyllidis, S. Bardenhagen, "On higher order gradient continuum theories in 1-D nonlinear elasticity. Derivation from and comparison to the corresponding discrete models", *Journal of Elasticity* **33**:3 (1993), 259-293.

[Vasiliev et al. 2002] A.A. Vasiliev, S.V. Dmitriev, Y. Ishibashi, T. Shigenari, "Elastic properties of a two-dimensional model of crystals containing particles with rotational degrees of freedom", *Physical Review B* **65** (2002), 094101.

[Vasiliev and Miroshnichenko 2005] A.A. Vasiliev, A.E. Miroshnichenko, "Multi-field modelling of Cosserat solids", *Journal of the Mechanical Behavior of Materials* **16**:6 (2005), 379-392.

[Vasiliev et al. 2005] A.A. Vasiliev, S.V. Dmitriev, A.E. Miroshnichenko, "Multi-field continuum theory for medium with microscopic rotations", *International Journal of Solids and Structures* **42**:24-25 (2005), 6245-6260.

ALEKSEY A. VASILIEV: aleksey.vasiliev@gmail.com
*Department of Mathematical Modelling, Tver State University,*
*Sadoviy per. 35, 170002 Tver, Russia*

ANDREY E. MIROSHNICHENKO: andrey.miroshnichenko@anu.edu.au
*Nonlinear Physics Centre, Research School of Physical Sciences and Engineering,*
*The Australian National University, Canberra ACT 0200, Australia*

MASSIMO RUZZENE: massimo.ruzzene@ae.gatech.edu
*School of Aerospace Engineering, Georgia Institute of Technology,*
*270 Ferst Drive, Atlanta GA, 30332, United States*